\documentclass{article}
\usepackage[style=philosophy-classic,shorthandintro=true,natbib=true]{biblatex}
\usepackage[english]{babel}
\usepackage{graphicx,csquotes,xcolor,epigraph,enumerate,amsmath,amsfonts,amssymb,forest,authblk}
\usepackage[margin=2cm]{geometry}
\usepackage[colorlinks,linkcolor=blue,urlcolor=blue,citecolor=black,plainpages=false,pdfpagelabels]{hyperref}

\title{\sc Realistic From Far But Far From Realism:\\Withering Scientific Realism in the Quantum Case}

\author[1]{\sc Raoni W. Arroyo\thanks{rwarroyo@unicamp.br}}
\affil[1]{Centre for Logic, Epistemology and the History of Science, University of Campinas, Campinas, SP, Brazil. Support: grant \#2021/11381-1, São Paulo Research Foundation (FAPESP).}

\author[2--5]{\sc Christian de Ronde}
\affil[2]{Philosophy Institute Dr. A. Korn, University of Buenos Aires, Buenos Aires, Argentina.}
\affil[3]{Fellow Researcher of the National Scientific and Technical Research Council (CONICET).}
\affil[4]{Institute of Engineering, National University Arturo Jauretche, Buenos Aires, Argentina.} 
\affil[5]{Center Leo Apostel for Interdisciplinary Studies, Brussels Free University, Brussels, Belgium.}

\date{ }

\begin{document}
\sloppy\raggedbottom
\maketitle

\begin{abstract}
\noindent Much has been discussed in the philosophy of science about how we should understand the scientific enterprise. On the one hand, scientific realists believe that empirically adequate theories can be supplemented by interpretations that can mirror \textit{reality-as-it-is}; on the other hand, anti-realists argue that this is not the case, as long as scientific theories make sufficiently accurate experimental predictions the addition of narratives is irrelevant for the scientific enterprise, and regarding narratives, it is preferable to remain agnostic. In this paper, we argue that realism was never really at stake in this debate.

\smallskip

\end{abstract}

\bigskip

\bigskip

\bigskip

\bigskip

\epigraph{``How many legs would a dog have, if you called his tail one?'' ``Five, of course.'' ``No; only four. It wouldn't make his tail a leg to call it one.''}{The Genesee Farmer}

\section*{Introduction}

Let's start by making the following statement: we don't know what Quantum Mechanics (QM) is telling us about the world in which we live. Nowadays this is almost a truism in the literature that deals with the philosophy of physics; yet, we feel that statements like this have persisted as commonsensical for far too long. And even though we observe an ever-growing number of so-called ``interpretations'' to cope with such a difficulty \citep[cf.][]{freirejr2022}, the task remains unsuccessful \citep[cf.][]{laloe2022}. With too many theoretical and interpretational options, and with the high hopes of singling out one which corresponds to reality, the search for the nexus between the theory and reality is still a work in progress. To be fair, more work than progress. Of course, this problem is deeply entrenched in the more general debate between realism and anti-realism in the philosophy of science. A debate marked by its polarization: the ontologically-oriented believers versus the empirical science-oriented down-to-earth skeptics. Using quantum mechanics as a background for a methodological discussion, in this article we argue that these problems are generated by some essential misunderstandings regarding the meaning of realism. In particular, we critically discuss if scientific realism can be actually considered as part of the realist program of science. Spoiler alert: it is not. 

The paper is structured as follows. In section \ref{sec:1} we set the stage with textbook definitions of anti-realism and scientific realism. Section \ref{sec:2} deals with how the measurement problem is understood in QM, and how scientific realists try to solve it through the addition of interpretations. Section \ref{sec:3} tries to disentangle the meaning of `theory' and `interpretation' within QM and clarify their role within scientific realism. Section \ref{sec:4} presents the problem(s) of underdetermination which plagues the realist understanding of QM. Sections \ref{sec:5} and \ref{sec:6} expose, respectively, the anti-realist's and the scientific realist's way out of underdetermination. Finally, section \ref{sec:7} discusses the death of realism, and weighs its impact.
It is worth mentioning that through this work we make heavy use of quotations, so we can name names and bring textual evidence to the current state of the art on the literature concerning both scientific realism and anti-realism.

\section{Scientific realism vs anti-realism}\label{sec:1}

According to contemporary common wisdom ``empirical science'' begins with observable data. Richard Feynman, maybe the most important physicists of the 20th century, explains this in the following terms:
\begin{quote}
In the beginning of the history of experimental observation, or any kind of observation on scientific things, it is intuition, which is really based on simple experience with everyday objects, that suggests reasonable explanations for things. But as we try to widen and make more consistent our description of what we see, as it gets wider and wider and we see a greater range of phenomena, the explanations become what we call laws instead of simple explanations. One odd characteristic is that they always seem to become more and more unreasonable and more and more intuitively far from obvious. \textelp{} There is no reason why we should expect things to be other- wise, because the things of everyday experience involve large numbers of particles, or involve things moving very slowly, or involve other conditions that are special and represent in fact a limited experience with nature. \citep[p.~127]{feynman1965}.
\end{quote}
Another influential U.S. physicist, Steven Weinberg, makes the same observational premise explicit in the following passage: 
\begin{quote}
\textelp{A}ll the sciences depend finally upon observation, psychological laws of behaviour as well as the physical axioms depend thereupon. What language system we employ, and what valid psychological principles we admit, will in some sense depend upon our observations. It cannot, therefore, be wholly a matter of decision or convention that certain physical axioms and psychological laws are adopted. \citep[p.~261]{weinberg2001}. 
\end{quote}
Of course, one can only understand the kernel role played by observation within contemporary physics when addressing the historical development of the discipline during the 20th century guided by Ernst Mach's positivist achievements. It is indubitable that his empiricist understanding of physics as ``an economy of experience'' would prove essential for the creation, during the 20th century, of both QM and relativity theory. Both Einstein and Heisenberg would follow Mach's observability principle when developing their theories. While Einstein would apply it in order to criticize the notion of simultaneity, \citet{heisenberg-1925} would develop his matrix mechanics as ``an attempt [...] to obtain bases for a quantum-theoretical mechanics based exclusively on relations between quantities observable in principle.'' Following this anti-realist trend of thought, Niels \citet[p.~10]{bohr1963} would conclude that ``[p]hysics is to be regarded not so much as the study of something a priori given, but rather as the development of methods of ordering and surveying human experience.'' 
But the influence of Mach did not restrict itself to physics. In the philosophical arena, logical positivists congregated in the \textit{Ernst Mach Society} would state in their famous manifesto:
\begin{quote}
Everything is accessible to man; and man is the measure of all things. Here is an affinity with the Sophists, not with the Platonists; with the Epicureans, not with the Pythagoreans; \textit{with all those who stand for earthly being and the here and now.} \citep*[p.~306, emphasis added]{manifesto1929}.
\end{quote}
In the contemporary literature the logical-positivist characterization of scientific theories, known as the \textit{syntactic approach}, continues to play an essential role within philosophical debates and even though it has been challenged by another approach called the \textit{semantic view}, the differences between them remain dubious \citep[cf.][]{lutz2015, halvorson2013}. While in the first a theory amounts to a ``partially interpreted abstract formalism, incorporating a language in terms of which the theory is formulated'' \citep[p.~3]{french2020}, the latter takes a theory to be the class of its models. In particular, as a defender of both anti-realism and the semantic view of theories, Bas \citet[p.~12]{vanfraassen1980} provided during the 1980s a clear definition of the role and scope of contemporary scientific theories: ``[s]cience aims to give us theories which are empirically adequate; and acceptance of a theory involves as belief only that it is empirically adequate.'' Thus, at this early stage, we might already conclude with Peter \citet[p.~9]{kosso2011} that: ``[a]ll scientific knowledge must be based on observation. It must have empirical foundations.'' 

However, even though the empiricist account of science was meant to replace the earlier reference to an underlying reality-in-itself, after its final triumph famously declared by Karl Popper in the late 1950s even anti-realists would become interested in re-introducing reality as a main goal of empirical science itself. From the ashes of realism, during the 1970s, Hilary Putnam and Richard Boyd would give birth to \textit{scientific realism}, characterized in terms of two essential commitments \citep[p.~179]{putnam1975}:
\begin{itemize}
{\bf \item[(1)] Science aims to give a literally true account of the world.
\item[(2)] To accept a theory is to believe it is (approximately) true.}
\end{itemize}
It is worth mentioning that ``truth'' here is being understood within the correspondence account of truth. Not every author is explicit about this, but some are, such as \citet[p.~96]{ladymanross2007} and \citet[p.~293]{ruetsche2020}. Arthur Fine expresses this idea as follows: 
\begin{quote}
The realist adopts a standard, model-theoretic, correspondence theory of truth; where the model is just the definite world structure posited by realisms and where correspondence is understood as a relation that reaches right out to touch the world. \citep[p.~137]{fine1986}.
\end{quote}
Roughly speaking, it is often ---and tacitly--- assumed by scientific realists that a theory is \textit{true} because it corresponds directly to the state of affairs in which the world actually \textit{is}. In this way they seem to confront the orthodox contemporary understanding of empirical science restricted to observational-predictive statements. Following then scientific realism, many contemporary philosophers of science argue in favor of the need to extend the goal of science beyond observability, through the re-introduction of ontological and/or metaphysical commitments. Yet, within this trend of thought, metaphysics is still understood as a discipline domesticated by science. For example, after conceiving that ``[m]etaphysics is ontology'', Tim Maudlin states that the ``the proper object of most metaphysics is the careful analysis of our best scientific theories''. Once again, the idea is grounded on an essentially empiricist standpoint:
\begin{quote}
\textelp{} the basic approach to ontology is always \textit{to start with the world as given} (the manifest image), and then adjust or adapt or modify or complicate that picture as needed, under the pressure of argument and observation. \citep[p.~104; p.~127, emphasis added]{maudlin2007}.
\end{quote}
The result of this modification to ``the manifest image'' is what \citet{sellars1962} has termed ``the scientific image'', a picture that escapes our commonsensical observations of the world that surround us and consequently becomes more complex and un-intuitive \citep[cf. also][]{vanfraassen1980}. Thus, the crossroad between the scientific realist and the anti-realist is set. Let's provide a textbook definition of both paths.
\begin{quote}
The basic idea of scientific realism is straightforward. Realists hold that science aims to provide a true description of the world, and that it often succeeds. So a good scientific theory, according to realists, is one that truly describes the way the world is. This may sound like a fairly innocuous doctrine. For surely no one thinks that science is aiming to produce a false description of the world? But that is not what anti-realists think. Rather, anti-realists hold that the aim of science is to find theories that are \textit{empirically adequate}, i.e. which correctly predict the results of experiment and observation. If a theory achieves perfect empirical adequacy, the further question of whether it truly describes the world is redundant, for anti-realists; indeed some argue that this question does not even make sense. \citep[p.~55, original emphasis]{okasha2016}.
\end{quote}

To sum up, while anti-realism presents theories as instrumental schemes which provide predictions about observations, the realist claims that science is committed to a descriptive account of nature which requires the introduction of metaphysical narratives capable of describing the underlying ``furniture of the world'' in a truthful correspondentist manner \citep{French2018}. Now, as remarked by \citet{vanfraassen1980}, even though anti-realists might accept that scientific theories can be understood in terms of such narratives, and that they could in principle be even \textit{true}, there is no need to \textit{believe} that such is the case. This marks the distinction between the scientific realist and the anti-realist. While the first grounds his interpretative claims exclusively in his faith, thus becoming some sort of ``fanatic-believer'', the latter seems to present a more down-to-earth rational position which van Fraassen has called ``agnosticism'' \citet{vanfraassen1980}. 

\section{The quantum case: (anti-realist) collapses and the (realist) measurement problem}\label{sec:2}

QM seems to be the perfect case study because even though it is a well-confirmed empirical theory, after more than a century since its development there seems to exist no consensus at all ---apart from the widespread claim that ``QM talks about microscopic particles''--- about what this theory tells us about the world in which we live \citep{mermin2014}. The failure to resolve this problem led \citet[p.~129]{feynman1965} to conclude already in the 1960s that ``[n]obody understands quantum mechanics''. This famous dictum has become not only an accepted standpoint for teaching the theory of quanta in Universities all around but also a justification for the common mantra with which Professors respond to inquisitive students: ``Shut up and calculate!'' Historically, as we mentioned above, it is by avoiding realist questions that QM was actually developed. As recalled by Arthur Fine:
\begin{quote}
\textelp{The} instrumentalist moves away from a realist construal of the emerging quantum theory were given particular force by Bohr's so-called philosophy of complementarity. This nonrealist position was consolidated at the time of the famous Solvay Conference, in October 1927, and is firmly in place today. Such quantum nonrealism is part of what every graduate physicist learns and practices. It is the conceptual backdrop to all the brilliant successes in atomic, nuclear, and particle physics over the past fifty years. Physicists have learned to think about their theory in a highly nonrealist way, and doing just that has brought about the most marvelous predictive success in the history of science. \citep[p.~124]{fine1986}.
\end{quote}
Such a triumph of instrumentalism within physics and its clear cut separation from philosophical issues is also exposed by Maudlin's very recent description of how, also in the 21st century, the so-called ``standard'' version of QM ---put forward by Bohr, Dirac and von Neumann--- remains nothing more than an instrumental ``recipe'': 
\begin{quote}
What is presented in the average physics textbook, what students learn and researchers use, turns out not to be a precise physical theory at all. It is rather a very effective and accurate recipe for making certain sorts of predictions. What physics students learn is how to use the recipe. For all practical purposes, when designing microchips and predicting the outcomes of experiments, this ability suffices. But if a physics student happens to be unsatisfied with just learning these mathematical techniques for making predictions and asks instead what the theory claims about the physical world, she or he is likely to be met with a canonical response: Shut up and calculate! \citep[pp.~2--3]{maudlin2019}. 
\end{quote} 
The contemporary division of labour imposes that while the physicist must concentrate on the pragmatic application of this ``recipe'' in order to produce technical developments, it is the exclusive task of philosophical research to discuss what the theory is really talking about.

The first closed and consistent mathematical formalism capable to account for quantum phenomena was famously proposed by Werner Heisenberg in July 1925. The German physicist had applied Mach's observability principle in order to escape the question of trajectories of elementary particles focusing instead in what was actually observed in the lab, namely, the intensive values of energy of line-spectra. The lack of a realist spatiotemporal image was of course one of the main obstacles which matrix mechanics needed to confront. But there would be no time to develop a meaningful answer. Just six months later, Erwin Schr\"odinger would present a wave equation that promised to restore the classical \textit{continuous} representation. As we all know, the lack of a truly consistent solution to this problem would then pave the road to the orthodox axiomatic version of QM which intended to find a common formal ground for both matrix and wave mechanics.\footnote{See \citet{derondemassri2022} for a more detailed formal historical account of the development of the theory.} In the year 1930, Paul Maurice Dirac would publish the first main treaty of what is still today known as Standard QM (SQM). Within this new formulation, given the conflict between on the one hand, the empirical-positivist understanding of physical theories in terms of (binary) observations, and on the other hand, the existence of \textit{quantum superpositions}, Dirac would famously introduce a measurement rule that would become to be known as ``the collapse of the quantum wave function''. Let us discuss this in some detail. SQM typically describes the states of physical systems as unit vectors which in Dirac's notation are written as $|\psi\rangle$. The dynamics of the theory is specified by Schr\"odinger's linear equation which, as a consequence of its linearity, imposes the appearance of quantum superpositions; i.e., linear combinations of other different vectors interpreted as \textit{states}. This feature of QM enables one to describe the state of a physical system as a superposition of other distinguishable states. As described in the orthodox literature, suppose a state of a quantum system that enters in an experimental setup has two possible outcomes: $|\psi_1\rangle$ and $|\psi_2\rangle$, with equal probability. The state can be then written as a superposition of these two different states $\alpha|\psi_1\rangle+\beta|\psi_2\rangle$, with $|\alpha|^2=|\beta|^2$.\footnote{Even though there might exist serious inconsistencies in some of these definitions \citep[see][sect.~4]{derondemassri2021}, we'll go with the flow of orthodoxy in presenting the measurement problem as such.} So far, so good. Now, according to \citeauthor{dirac1930}'s famous 1930 treatise, the quantum mechanical formalism describes quantum systems as superpositions but we never observe them, instead, we observe single (binary) ``clicks'' in detectors \citep[cf. also][]{vonneumann1932}. Now, given that according to Dirac \citet[p.~3]{dirac1930} ``\textelp{} science is concerned only with observable things \textelp{}'' and that ``whether a picture exists of not is a matter of only secondary importance'' \citep[p.~10]{dirac1930} the problem becomes then: how do we relate quantum superpositions (e.g. $\alpha|\psi_1\rangle+\beta|\psi_2\rangle$) with the single outcomes we actually observe in the lab (e.g., \textit{either} $|\psi_1\rangle$ or $|\psi_2\rangle$)? In order to ``bridge the gap'' Dirac would introduce an \textit{ad hoc} rule according to which when measuring a quantum superposition the observation would produce a ``jump'' to the measurement outcome. From an anti-realist perspective this is not problematic at all. After all, following an instrumentalist account of the theory, one could refuse to reify what happens in a measurement process as long as ``the calculations are right'' \citep{mermin2014, fuchsperes2000}. However, from a realist viewpoint things are quite different. The \textit{ad hoc} addition of this ``collapse'' ought to be explained. Thus, bluntly put, the standard way of presenting the measurement problem is this: how can we explain the ``collapse'' that takes place when measuring quantum superpositions? As we mentioned above this is not considered to be a task to be performed by physicists but one exclusively linked to philosophers of physics. According to them what is missing is an ``interpretation'' of the theory which explains what is actually going on in the measurement process \citep{Jammer1974, ruetsche2002, callender2020}. Thus, trying to solve the so-called ``measurement problem'' of quantum mechanics ---also traditionally called the ``problem of the collapse'' \citep[cf.][]{pessoajr2022}--- implies the addition of a narrative that would explain ``what is QM really talking about'' \cite{mermin1998what}. As \citet{maudlin2011} stresses: 
\begin{quote}
The most pressing problem today is the same as ever it was: to clearly articulate the exact physical content of all proposed `interpretations' of the quantum formalism is commonly called the measurement problem, although \textelp{} it is rather a ``reality problem''.
\citep[p.~52]{maudlin2011}.
\end{quote}

A clear way of posing the problem is the following: given a specific basis (or context), QM describes mathematically a quantum state in terms of a superposition of, in general, multiple states. Since the evolution described by QM allows us to predict that the quantum system will get entangled with the apparatus and thus its pointer positions will also become a superposition, the question is why do we observe a single outcome instead of a superposition of them? More precisely: given a quantum system represented by a superposition of more than one term, $\sum c_i | \alpha_i \rangle$, when in contact with an apparatus ready to measure, $|R_0 \rangle$, QM predicts that system and apparatus will become ``entangled'' in such a way that the final ``system + apparatus'' will be described by $\sum c_i | \alpha_i \rangle |R_i \rangle$. Thus, as a consequence of the quantum evolution, the pointers have also become ---like the original quantum system--- a superposition of pointers $\sum c_i |R_i \rangle$. Another way to present the measurement problem has become popular:\footnote{Examples of the use of this way of treating the measurement problem as numerous. See, for instance, \citet{ladymanross2007, esfeld2019, friederich2014, allori2021}.} \textit{Maudlin's trilemma}, is defined as the problematic conjunction of the three assumptions made upon a quantum-mechanical description (often presented as the evolution of the wave function):
\begin{quote}
1.A The wave-function of a system is complete, i.e. the wave-function specifies (directly or indirectly) all of the physical properties of a system.\\1.B The wave-function always evolves in accord with a linear dynamical equation (e.g. the Schr\"odinger equation).\\1.C Measurements of, e.g., the spin of an electron always (or at least usually) have determinate outcomes, i.e., at the end of the measurement the measuring device is either in a state which indicates spin up (and not down) or spin down (and not up). \citep[p.~7]{maudlin1995}.
\end{quote}

Independently of its different expositions, philosophers of QM agree that the way to solve the measurement problem is by adding ---what philosophers call--- an ``interpretation''.\footnote{It is worth mentioning that, by selecting such literature, we are not dealing with the so-called ``hermeneutic'' literature that traditionally deals with ``interpretation'' in philosophy; nevertheless, to be fair, \citet{muller2015} did exactly such incursion and showed that the notion of ``interpretation'' in the philosophy of physics is sui generis, viz., the one(s) that we are discussing in this article.} This is a term which is not known by physicists,\footnote{In fact, as \citet[p.~59]{schlosshauer2011} has described: ``[i]t is no secret that a shut-up-and-calculate mentality pervades classrooms everywhere. How many physics students will ever hear their professor mention that there's such a queer thing as different interpretations of the very theory they're learning about? I have no representative data to answer this question, but I suspect the percentage of such students would hardly exceed the single-digit range.''} so let us see what philosophers exactly mean by it.

\section{Saving phenomena or interpreting reality (...or both)? }\label{sec:3}

According to \citet{saatsi2019}, the ``standard story'' of scientific realists meeting quantum mechanics goes like this.
\begin{quote}
The standard (`textbook') QM, which incorporates the collapse postulate, is not amenable to a realist attitude towards the dynamics of the theory, given the irreducible role played by the notion of measurement as yielding determinate observable measurement outcomes. The upshot, then, is that the orthodox QM, unvarnished with a `realist interpretation', is best regarded as a mere instrument or recipe for making predictions. Avoiding such blatant antirealism about QM ---as the realist desires--- thus requires articulating and defending a variant of QM that does not involve the problematic collapse postulate inconsistent with the unitary quantum dynamics. That is, it requires articulating and defending a variant of QM amenable to a realist interpretation. \citep[p.~157]{saatsi2019}.
\end{quote}
But what is a ``realist interpretation''? 

To \citet[p.~226]{vanfraassen1989} ``any question about content'' of a scientific theory is met with an interpretation. Such content, says \citet[p.~242]{vanfraassen1991}, is ``\textit{the question of interpretation:} Under what conditions is this theory true? What does it say the world is like? These two questions are the same''. This definition of interpretation is often used as a basis for more contemporary ones. \citet[p.~293]{ruetsche2018} goes further, stating that the \textit{realist content} of scientific realist that believes in a theory $T$ is the content of an \textit{interpretation} of $T$, viz. an account of how can the world possibly be \textit{if} $T$ is true. Pushing the notion of ``interpretation'' even further, \citet[p.~212, original emphasis]{williams2019} states that ``the goal of interpreting physical theory is to identify and characterize its \textit{fundamental} ontological structure''. This suggests that there is a strict link between theories' interpretation and ontology (more on that later). Things are quite complicated for even though most physicists prefer to avoid entering the debate about reference, there is anyhow a generalized consensus that SQM ---what Maudlin calls a ``recipe''--- describes a microscopic world composed by quantum particles. This is obviously an ontological claim. So it seems, according to some physicists, the ontological level might seem part of physics. This situation, however, leaves us with a deeper question: what is a physical theory, after all? Is it only meant to ``save phenomena'' or to actually describe the entities that compose reality? Because if physical theories do contain an ontological level within then anti-realists are essentially wrong: theories do not merely talk about observations, they also describe reality. But if ontology is already included within theories by the physicist then what is the role of philosophers? Are they supposed to add an ontology to already empirical adequate theories or do they have to explain what physicists already mention when talking about their own theories? For example, Maudlin \citep[p.~4, original emphasis]{maudlin2019}, as a philosopher claims that the ``quantum recipe'' is not a theory because it lacks an ontology. ``A physical theory should contain a physical \textit{ontology}: What the theory postulates to exist as physically real. And it should also contain \textit{dynamics}: laws (either deterministic or probabilistic) describing how these physically real entities behave.'' But Maudlin also claims that it is the task of philosophers to add this ontological level. So do physical theories require philosophy? Or is philosophy part of physics? Should we take into account what physicists say about quantum particles or is this ``just a way of talking''?\footnote{Even though physicists agree that QM talks about ``quantum particles'', there is no consensus about what such particles are meant to be \citep[cf.][]{wolchover2020}.} This level of confusion and inconsistency threatens rational debate. However, things become much more clear when we look at the division of labor. In this case, it is absolutely clear that any research program related to the problem of reality must be undertaken by philosophers, not by physicists. This is particularly clear in the context of QM. As \citet{Carroll20} has recently described: ``[m]any people are bothered when they are students and they first hear [about QM]. And when they ask questions they are told to shut up. And if they keep asking they are asked to leave the field of physics.''

The standard philosophical account of realism implies that ``the stuff the world is made of'' is essentially composed of ``things'' or ``entities''. It is theories and models which describe them, firstly, in terms of mathematical representations, and more specifically, through the addition of ontological interpretations. At first sight it is only for realists that the ontological content of a theory is of crucial importance \citet[p.~2]{durrlazarovici2020}: ``\textelp{} the ontology of a physical theory specifies what the theory is about''. One way to investigate this problem is to check the ontological commitments of theories \textit{\`a la} \citet[p.~65]{quine1951}, i.e., to investigate ``what, according to that theory, there is''. It is here, however, where the conflation between ``theory'' and ``interpretation'' starts to kick in.\footnote{To complicate things, anti-realists have also called pragmatic schemes ``interpretations'' \citep[e.g. QBism, Copenhaguen, modal ``interpretations'', cf.][]{fuchsperes2000, vanfraassen1991,cushing1994}.} To \citet{durrlazarovici2020}, the term is inadequate.
\begin{quote}
A poem is interpreted if you want to elicit some deeper meaning from the allegorical language. However, physical theories are not formulated in allegories, but with precise mathematical laws, and these are not interpreted, but analysed. So the goal of physics must be to formulate theories that are so clear and precise that any form of interpretation ---what was the author trying to say there?--- is superfluous. \citep[p.~viii]{durrlazarovici2020}.
\end{quote}

As it stands there seem to exist two main possibilities to solve the measurement problem. Either we introduce a convincing narrative to SQM that explains what is really going on in the measurement process or we must change the theory itself. This can be done by either erasing the collapse from the formalism (as in Bohmian mechanics) or explaining it as a new physical process (as in collapse theories). Today's most popular ``realist interpretations'' ---at least, according to some of the most influential philosophers of physics--- are good examples of these popular strategies. Assuming that the quantum formalism is not complete and the collapse is unnecessary, hidden variable theories such as Bohmian mechanics \citep{bohm1952} have been proposed; assuming that the collapse is a real process theories with additional stochastic dynamics have been also suggested \citep*[e.g.][]{GRW}; also, interpreting SQM in a third family of so-called ``non-collapse interpretations'' we find many-worlds interpretations \citep{everett1957, wallace2012}.\footnote{As it has been remarked by \citet{barrett2011} and \citet{conroy2012}, however, in Everett's own manuscript \citep{everett1957} there is no metaphysical commitment nor reference to the existence of many worlds. On the very contrary, the popularization of Everett's work in many worlds was due to \citet{dewitt1971}. Nevertheless, we acknowledge the existence of what can be called an ``Everettian orthodoxy'' which links Everettian quantum mechanics with many worlds \citep[cf.][]{wallace2012}.} This is today's standard taxonomy for so-called ``interpretations of QM'', defined as solutions to the measurement problem. \citet[p.~7]{maudlin1995} himself believes that ``[a]ny real solution [to the measurement problem] demands new physics'', so this is where things become tricky. Notice that the literature has grown in considering that realist approaches to QM must solve the measurement problem and this is done by offering an \textit{interpretation}. For instance, it is not uncommon to read statements such as these in the literature.
\begin{quote}As is well known among philosophers of physics, nonrelativistic quantum mechanics permits of many distinct formulations that are regarded as serious rivals by cross-sections of the physics community. For example, two alternative formulations of nonrelativistic quantum mechanics (or perhaps, classes of formulations) are the Everettian (many worlds) theory and Bohmian mechanics. \citep[p.~68]{ney2012}.\end{quote}

The first thing to acknowledge is the difficulty to disentangle ``theory'' and ``interpretation''. Of course, not everyone bothers with such distinction. Some acknowledge the differences between ``theory'' and ``interpretation'' but leave aside the concerns of explaining it \citep[cf.][p.~78, fn.~3]{french2013}. To \citet[p.~294]{ruetsche2018}, ``[w]hether to count the Bohmian answer as an interpretation of QM is an idle bookkeeping matter''. Such a distinction plays a kernel role in the realist-antirealist debate and must not be taken for granted. It is unanimously accepted that a \textit{theory} is a mathematical formalism/model capable of providing predictions about future empirical observations \citep[cf.][]{Jammer1974}. This is implied within the syntactic view of scientific theories \citep{lutz2012}, where the division between `theoretical terms' and `observational terms' is bridged via empirical \textit{correspondence rules}, namely, an \textit{empirical interpretation}. However, such interpretational rules do not provide a narrative about reality but are only added to ``save the phenomena''.\footnote{This is another example of the way a term like ``interpretation'' is used in different ways.} So physical theories imply an already empirically interpreted theory right from the start ---otherwise it is not physics, but mathematics. Here's \citet{cassini2016} on that matter:
\begin{quote}
Quantum mechanics, like any physical theory, is a formalism that already has an interpretation in physical terms. \textelp{} Without a minimal interpretation, quantum theory could not be considered as a physical theory, since it would be reduced to a pure mathematical formalism without empirical content. \citep[pp.~22--25]{cassini2016}.
\end{quote}
As we mentioned above, at this point the empirical theory still lacks a description of what the world is like. The \textit{interpretational rules} ---such as the measurement postulate of QM--- are only there in order to ``save the phenomena''. Once again, many physicists and philosophers claim that a theory should also provide an ontological narrative that tells us how the world would be if the theory was true. In this respect, \citet[p.~281]{sklar2003} has already remarked his hesitation on these matters: ``I doubt that one can draw any principled line between replacing a theory and `merely interpreting' it''. Indeed, there is difficulty in doing so \textit{mainly} because (at least in the quantum case) it is not clear what counts as the ``uninterpreted theory'' which is ``interpreted'' through the many narratives added to SQM.\footnote{To cope with such a difficulty, \citet{muller2015} proposed a ``QM$_0$'', and \citet{arroyoolegario2021} proposed a ``QM$^{bas}$''. Other formulations with the same minimalist scope can be found in recent literature \citep[cf.][]{wallace2020, acuna2021}.} In fact, SQM is many times referred to by physicists as ``the Copenhagen interpretation''. Furthermore, very often there is a mix between the empirical interpretation and the ontological one. Often different physical theories are intertwined with interpretations in an empirical sense because, as famously stated by \citet[p.~237]{carnap1966}, ``[a] postulate system in physics cannot have, as mathematical theories have, a splendid isolation from the world''. The problem begins to show itself when we ask the following question: what does ``the world'' mean here? Is it referring to empirical observations or to a description of the entities that inhabit reality? In this respect, Bohmian mechanics, even though possesses many different interpretations come right from the start with a story about a world where particles behave in a non-local manner directed by a quantum field. So is this a theory or an interpretation?
\begin{quote}
Throughout the scientific and philosophical literature on hidden variables, Bohm's theory is called, interchangeably, ``interpretation'', ``theory'', ``model'', ``approach'', ``point of view'', ``formulation'', ``version'', ``explanation'' and ``alternative'', among other expressions. \citep[pp.~14--15]{cassini2016}.
\end{quote} 
In order to solve this problem, we'll be using as a rule of thumb the following demarcation: the theory is the mathematical formalism plus the empirical rules that relate them to observations. Following orthodoxy, that is after all what ``empirical science'' aims to do, namely, to develop ``empirical adequate theories''. So once the formalism is changed, the theory has changed as well. But once the theory has changed, the range of possible ontological interpretations is also changed:
\begin{quote}
\textelp{} whatever the conditions that an interpretation of a physical theory, such as the standard quantum theory, must satisfy, the interpretation cannot consist in reformulating the theory, for example, by another formalism or by means of other axioms or postulates, nor can it consist in replacing this theory for an alternative theory. \citep[p.~15]{cassini2016}.
\end{quote}
In the standard taxonomy, empirical theories ($\mathfrak{Emp}_{T}$) are mathematical formalisms that are already interpreted with empirical rules \citep{cassini2016, muller2015, arroyoolegario2021}. One of the best examples is of course the ``standard recipe'' of QM which is clearly not ``just mathematics'', i.e. it is not just vectorial algebra. In fact, many mathematicians take courses on vectorial algebra making no mention of quantum states, the projection postulate or measurement results. A pure mathematical formalism cannot predict actual observations or ``clicks'' in detectors simply because it refers to nothing but abstract mathematics, it has no conceptual content that can be read literally from it. In fact, the claim that the formalism of QM has a ``literal reading'' \citep[e.g.,][]{wallace2012} is \textit{de facto} contradicted by the actual existence of an enormous number of narratives. Of course, one can also present different empirical theories which account for exactly the same set of observations. In the case of QM, we have indeed, different empirically adequate mathematical formalisms. 
\begin{table}[ht]
\centering
\begin{tabular}{ll}
\hline\noalign{\smallskip}
\textbf{$\mathfrak{Emp}_{T}$}\\ 
\noalign{\smallskip}\hline\noalign{\smallskip}
Standard Quantum Mechanics \\ \noalign{\medskip}\noalign{\smallskip}
Dynamical Reduction Program \\ \noalign{\medskip}\noalign{\smallskip}
Bohmian Mechanics \\ \noalign{\smallskip}\hline
\end{tabular}
\label{tab:interpretation-emp}
\caption{Empirical theories.}
\end{table}
One might choose to provide an ontological account of SQM but one might prefer to interpret Bohmian Mechanics or the Dynamical Reduction Program.\footnote{Regarding Bohmian mechanics, the reader should be referred to \citet{sole2009phd, sole2012, sole2017}; for the Dynamical Reduction Program (also known as ``Collapse Theories''), see \citet{ghirardi-bassi2020,CSL1,CSL2,GRW}.} As we already mentioned, $\mathfrak{Emp}_{T}$ gives us an \textit{empirically adequate theory}, viz. a mathematical model which accounts for observations in the lab. As far as anti-realism ---pace logical empiricism--- goes, this is as good as it gets: it provides the correspondence rules which enable one to connect the abstract theoretical terms of a mathematical formalism to an empirically adequate vocabulary, viz. to an observational vocabulary (e.g. protocol sentences). For instance, here's a common statement of what is an interpretation of QM:
\begin{quote}
 What must an interpretation of quantum mechanics do in order to be considered viable? We suggest that one vitally important criterion is the following: \textit{Any successful interpretation of quantum mechanics must explain how our empirical evidence allows us to come to know about quantum mechanics.} That is, an interpretation of quantum mechanics must be able to tell a sensible story about how empirical confirmation works in the context of quantum-mechanical experiments \textelp{}. \citep[p.~1, original emphasis]{adlam2022}.
\end{quote}
This implies of course, as we already mentioned, a positivist way of framing scientific theories \citep[cf.][]{carnap1966,maxwell1962,suppe1977}. Both syntactic and semantic approaches to scientific theories follow these standpoints \citep[for a discussion on this topic, cf.][]{arroyoolegario2021,french2020,krausearenhart2016}. To \citet[p.~12]{vanfraassen1980}, the essential anti-realist claim is that ``[s]cience aims to give us theories which are empirically adequate; and acceptance of theories involves as belief only that it is empirically adequate''. Scientific realists, on the other hand, are more demanding people claiming that: ``[s]cience aims to give us, in its theories, a literally true \textit{story} of what the world is like; and acceptance of a scientific theory involves the belief that it is true'' \citep[p.~8, emphasis added]{vanfraassen1980}. As we have discussed above, this involves adding an ``ontological interpretation''; viz. a story concerning ``what is the world like if the theory was true'', or more specifically, an account of ``what is the kind of stuff that the theory refers to''. 

\begin{table}[ht]
\label{tab:interpretation-ont1}
\centering
\begin{tabular}{ll}
\hline\noalign{\smallskip}
\textbf{$\mathfrak{Emp}_{T}$} & \textbf{$\mathfrak{Int}_{ont}$} \\ \noalign{\smallskip}\hline\noalign{\smallskip}
Standard Quantum Mechanics & \begin{tabular}[T]{@{}l@{}}Ithaca Interpretation\\ Popper's Propensity Interpretation \\Diek's Modal Interpretation\\ Consistent Histories\\ Many-Minds Interpretation\\ Many-Worlds Interpretation\\ Heisenberg's Potentiality Interpretation \\ \end{tabular} \\\noalign{\medskip} \hline
\end{tabular}
\caption{Interpreting SQM in Realist Terms.}
\end{table}

\begin{table}[ht]
\label{tab:interpretation-ont2}
\centering
\begin{tabular}{ll}
\hline\noalign{\smallskip}
\textbf{$\mathfrak{Emp}_{T}$} & \textbf{$\mathfrak{Int}_{ont}$} \\ \noalign{\smallskip}\hline\noalign{\smallskip}
Bohmian Mechanics & \begin{tabular}[T]{@{}l@{}}Bohm's Interpretation\\ Healey \\ Flash \\ \end{tabular} \\\noalign{\medskip} \hline
\end{tabular}
\caption{Interpreting Bohmian Mechanics in Realist Terms.}
\end{table}

Tables 2 and 3 will help us to visualize this \textit{further} level of description. Given one accepts (say) SQM, one might introduce an ontological interpretation $\mathfrak{Int}_{ont}$ that describes what SQM really talks about. Of course, one might also choose to provide ontological interpretations for, say, Bohmian mechanics. To conclude, from a logical-empiricist-based account of theories, one might add a realist level of description through the introduction of an \textit{ontological interpretation} linked to reality-in-itself. It is at this point that things become even more complex and weird because many anti-realists accept the need to discuss the measurement problem but not in realist terms! Instead, anti-realists have entered the supposedly realist debate about reference by adding what might be called: ``non-ontological interpretations'' ($\mathfrak{Int}_{\neg ont}$). It becomes then very unclear what would be the difference between $\mathfrak{Emp}_{T}$ and $\mathfrak{Int}_{\neg ont}$? Or, in other words, what is the essential addition provided by such non-ontological interpretations? In short, what is the difference between $\mathfrak{Emp}_{T}$ and $\langle\mathfrak{Emp}_{T}+\mathfrak{Int}_{\neg ont}\rangle$? Why bother in adding an $\mathfrak{Int}_{\neg ont}$ if $\mathfrak{Emp}_{T}$ already provides everything which is expected from a scientific theory in terms of empirical adequacy?

\begin{table}[ht]
\centering
\begin{tabular}{ll}
\hline\noalign{\smallskip}
\textbf{$\mathfrak{Emp}_{T}$} & \textbf{$\mathfrak{Int}_{\neg ont}$} \\ \noalign{\smallskip}\hline\noalign{\smallskip}
Standard Quantum Mechanics & \begin{tabular}[T]{@{}l@{}}Copenhagen Interpretation\\ Everett Interpretation \\ van Fraassen's Modal Interpretation \\ Fuchs and Peres's ``No Interpretation'' \\ \end{tabular} \\\noalign{\medskip} \hline
\end{tabular}
\label{tab:interpretation-nonont}
\caption{Interpreting SQM in Anti-Realist Terms.}
\end{table}
\noindent Notice that we follow the distinction made by \citet{barrett2011} between the many worlds interpretation proposed by DeWitt which implies the real existence of a multiverse and the Everett interpretation which as an essentially positivist anti-metaphysical account of QM does not.

As we shall see in the following section, there are much deeper problems involved in the realist attempt to interpret a theory, independently of providing a clear definition of what a theory amounts to or what the role of interpretation really is. 

\section{Underdetermination and the ``map of madness''}\label{sec:4}

The underdetermination of theory by data is a very familiar topic in the general philosophy of science. It is a concern for scientific realism that arises from the fact that, in principle, there can be different scientific theories, $\mathfrak{Emp}_{T}$, which account for the same observational data \citep[cf.][]{vanfraassen1980}.\footnote{Historically this type of underdetermination was considered merely hypothetical or modal \citep[cf.][]{stanford2021}, but nowadays it is well established that quantum mechanics exemplifies well \citep[cf.][]{callender2020}, viz. with the solutions to the measurement problem \citep[cf.][]{maudlin1995, arroyoolegario2021,acuna2021} ---although there is a recent dissidence about that \citep[cf.][]{wallace2022}.} So how to choose one option between the many? This is known in the literature as the problem of 
underdetermination of theory by data. The same set of empirical observations can be predicted by different empirically equivalent mathematical formalisms. However, this problem is not restricted to observations. An analogous situation happens at the level of ontological interpretations, where we can also a find different narratives, $\mathfrak{Int}_{ont}$ to account for the same empirical theory, $\mathfrak{Emp}_{T}$. Consequently: as explained by \citet[p.~1123]{sklar2010}, ontological underdetermination implies that the furniture of our world can be understood in radically incompatible ways:
\begin{quote}
Our foundational theories usually exist in a scientific framework in which they are subject to multiple, apparently incompatible, interpretations. And given the interpretation you pick, your view of what the theory is telling us about the basic structure of the world can be radically unlike that of someone who opts for a different interpretation of the theory. \citep[p.~1123]{sklar2010}.
\end{quote}

In the last section we tried to disentangle foundational issues concerning the meaning of ``theory'' and ``interpretation''. With that distinction under our belts, recall that the notions of ``theory'' and ``interpretation'' are being employed loosely in the philosophy of physics. So one can wonder what is a theory, and how can a mathematical formalism talk about entities. However, even if we grant that by ``theory'' they mean $\mathfrak{Emp}_{T}$, this does not allow one to make any reference to entities in ontological terms for, as we discussed above, a scientific theory makes exclusively reference to observable events. In order to talk about entities, one needs to add the $\mathfrak{Int}_{ont}$ layer to $\mathfrak{Emp}_{T}$. Thus, the scientific realist is in trouble because she needs to bridge the gap, not only between empirical data and theories but also between theories and ontologies. Or in other words, scientific realists need to provide a rational theory-choice criteria to justify objectively the choice of a single $\langle\mathfrak{Emp}_{T}+\mathfrak{Int}_{ont}\rangle$; i.e. they need to justify why pick one single theory and one single ontological interpretation in amidst of a plethora of underdetermined options. 

But the problems for scientific realists do not finish here because, as Steven \citet{french2011} has remarked, we can also add the level of metaphysical underdetermination. Which specific metaphysical system is required for the specific choice of the chosen theory already interpreted ontologically? E.g. bundle theories \textit{vs} substratum theories of individuality. Regarding this last level of underdetermination, \citet{benovsky2016} proposes to appeal to aesthetic virtues:
\begin{quote}
\textelp{} the aesthetic properties of a theory can be appealed to when it comes to preferring one theory over another. In short, the view at hand is that metaphysical theories are beautiful and that contemplating their beauty is what drives us to prefer one to another. \citep[p.~vii]{benovsky2016}.
\end{quote}
In a same vein, but with a self-avowed realist guise, \citet[p.~215]{chakravartty2017} endorses voluntarism for similar cases of metaphysical disputes, i.e. the thesis according to which ``\textelp{} the relevant beliefs and actions are freely chosen, or voluntary, as opposed to being forced in virtue of reason alone''. Of course, if the same criteria is applied for one to choose between different $\langle\mathfrak{Emp}_{T}+\mathfrak{Int}_{ont}\rangle$, then the rational debate for theory choice in scientific realism is long gone. Unfortunately, this seems to be precisely the case. \citet[p.~231]{durrlazarovici2020}, after stating that ``\textelp{the} quest for an understanding of quantum mechanics lead us to three competing formulations rather than just one `final' version'', and after stating their preference for Bohmian mechanics, they conclude the following.
\begin{quote}
In science, we are always in the situation that observable phenomena can merely constrain but not conclusively determine the ``correct'' theory. When it comes to quantum physics, this is aggravated by the fact that our epistemic access to the microscopic state of affairs is limited, in principle, and we have striking illustrations of the fact that different theories can provide radically different descriptions of the world even when they are empirically equivalent. This is why we have to resort to criteria such as beauty, simplicity, and explanatory power, criteria that have a subjective quality and need not lead to a consensus. \citep[pp.~231--232]{durrlazarovici2020}.
\end{quote}

Due to underdetermination, the explanatory power of any theory supplemented by a narrative ($\mathfrak{Emp}_{T}+\mathfrak{Int}_{ont}$) are essentially equivalent. On the other hand, criteria such as beauty and simplicity concern \textit{us}, not nature. This is why \citet{maudlin2019interview} recognizes that: 
\begin{quote}
Different approaches would give you different answers [to the question of physical reality in QM]. My preference, my aesthetic sense is that the pilot wave seems more natural to me and the objective collapse seems to me more unnatural but I wouldn't ---you know--- give a lot of attention to my aesthetic preferences here. \citep{maudlin2019interview}.
\end{quote}

As a consequence, the scientific realist is left with the following set of problems: 
\begin{description}
\item[Theoretical underdetermination:] There are many different mathematical models capable to account for the same table of observable data. So how can we choose between them? E.g. Bohmian mechanics, Dynamical Reduction Program, or SQM \citep[cf.][]{durrlazarovici2020}?
\item[Ontological Underdetermination:] There are many different ontological interpretations that explain, in different ways, what are the entities of an interpretation of a theory. So how can we choose between them? E.g. within SQM, many minds or many worlds \citep[cf.][]{arroyoarenhart2022}?
\item[Metaphysical Underdetermination:] There are many different metaphysics that can account for the ontology of an interpretation of a theory. So how can we choose between them? E.g., within SQM and many worlds, individuals or non-individuals \citep[cf.][]{frenchkrause2006, french2011}? 
\end{description}

\begin{figure}[ht]
\centering
\begin{forest} for tree={grow=90}
[Observable Quantum Phenomena
[Bohmian QM
[Pilot wave ]
[Quantum potential ] 
]
[Dynamical Reduction Program
[Flashes]
[Dispositions] 
]
[Standard QM
[Potentialities]
[Propensities]
[Many worlds] 
]
]
\end{forest}
\caption{``A map of madness''}
\label{fig:mapofmadness}
\end{figure}
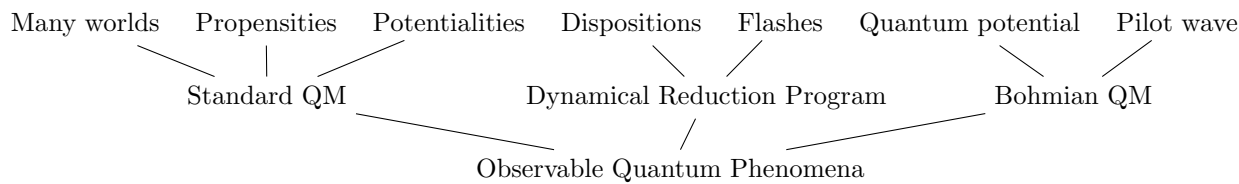

Let us use Figure \ref{fig:mapofmadness} as a means to be graphic about the underdetermination argument. In the first line we have the observational data. In the second line we have the $\mathfrak{Emp}_{T}$ surveyed in this paper. The third contains a few $\mathfrak{Int}_{ont}$ associated with each $\mathfrak{Emp}_{T}$ ---of course, there are many more in market. There is a fourth and final level we did not include where different metaphysical readings could be chosen to account for a particular ontological narrative. Of course this is an example with very few options yet good enough for a good old underdetermination argument. 

\noindent Today, the number of interpretations of QM continues to grow exponentially. As David \citet[p.~8]{mermin2012} described: ``[q]uantum theory is the most useful and powerful theory physicists have ever devised. Yet today, nearly 90 years after its formulation, disagreement about the meaning of the theory is stronger than ever. New interpretations appear every day. None ever disappear.'' This unconstrained reproduction of interpretations has lead the field into an untenable situation which has been described by Ádan Cabello as ``a map of madness''. A map which, we might add, floats free from science. Indeed, such ``additions'' are not considered by physicists as part of the scientific program. In this respect, Roberto \citet[p.~367]{torretti1999} is correct to point out that interpretations of QM must be considered only as ``meta-physical ventures \textelp{} for they view the meaning and scope of QM from standpoints outside empirical science.'' Thus, to conclude, we might say that it is quite clear that the complete lack of an objective set of rational scientific conditions that would allow us to select between the many options has turned this purely philosophical debate into a sterile un-scientific dispute. 

To sum up, there are no objective reasons that would allow us to choose one option between the many that appear in each level of underdetermination. And in the end, realists evaluate all these choices ---of formalisms, interpretations, ontologies and metaphysics--- only subjectively, in terms of aesthetic standards \citep{benovsky2016}, voluntarism \citep{chakravartty2017} or ---simply put--- in terms of ``personal preferences'' \citep{maudlin2019interview}. Obviously, these choices are exactly the opposite of a scientific objective procedure. However, and this is the essential point we want to make, say we were able to bypass theoretical underdetermination ---we do not know how--- and choose a mathematical formalism, and say we agree that an ontological interpretation is needed and we end up choosing one particular narrative ---we do not know how--- and, finally, say we choose a metaphysical reading of the ontology. Notice: this is still an \textit{empiricist} notion of interpretation right from the start ---recall, it is built on van Fraassen's definition! So when realists accept that the only thing we need to do in order to overcome the anti-realist understanding of science is to ``add narratives'', they're like the Trojans welcoming a dangerous wooden horse inside their walls.

\section{The anti-realist solution to underdetermination}\label{sec:5}

Now, there is a sense in which different theories can be welcomed by anti-realists. Take van Fraassen's constructive empiricism and its axiological statement regarding the aim of science:
\begin{quote}
\textelp{} constructive empiricism holds that the aim of science is not to find true theories, but only theories that are empirically adequate: that is, theories whose claims about observable phenomena are all true. Since the empirical adequacy of a theory is not threatened by the existence of another that is empirically equivalent to it, fulfilling this aim has nothing to fear from the possibility of such empirical equivalents. \citep[\S 3.2]{stanford2021}.
\end{quote}
As long as underdetermination doesn't threaten empirical adequacy, the underdetermined theoretical options are all good to go ---they're on the same epistemic level, so why bother choosing \textit{one}? But what about the multiplication of ontological interpretations and the map of madness it has delivered? \citet{bueno2021} has attempted to defend this situation by relating constructive empiricism to neo-Pyrrhonism. 
\begin{quote}In understanding critically each conception, the Pyrrhonist also understands something about the world if such a proposal correctly represented the way things are. Given that eventually Pyrrhonists suspend judgment about such proposals, they are not committed to their truth, but can perfectly coherently concede that the critical exploration of dogmatic proposals offers some understanding ---of the way things \textit{could be}, even though it is unclear that we can know how they actually are. (Ultimately, the neo-Pyrrhonist suspends judgment about such knowledge claims.) \citep[p.~13]{bueno2021}.
\end{quote}
According to Bueno, the growing number of interpretations (in our terminology, both $\mathfrak{Int}_{ont}$ and $\mathfrak{Int}_{\neg ont}$, although he doesn't distinguish between them ---nor anyone does, to our best knowledge) should be all welcome because such multiplicity enhances our (modal) \textit{understanding} of how the world could possibly be:
\begin{quote}
Although no commitment to the truth of any given interpretation of quantum theory is ever advanced, each interpretation is taken as providing some understanding. By examining the details of each interpretation, we understand various aspects of quantum mechanical objects and processes: how they emerge, which features they display, and why they behave the way they do ---in light of each interpretation. What we have is a form of pluralism, in which each interpretation contributes with a particular account of the overall picture, by indicating how things could be. \citep[p.~13]{bueno2021}.
\end{quote}
The idea is that the different interpretations provide an account of ``how the world could \textit{possibly} be'', without commitment in their truth. Embarrassingly enough, the above-mentioned anti-realist position is almost indistinguishable from a self-avowed realist position, such as Chang's:
\begin{quote}
    Give me the Heisenberg, Schrödinger, Feynman, \textit{and} Bohm versions of quantum mechanics: so many different ways of appreciating the physical world, more windows on nature, more enriched understanding of it. Why is this any worse (or better) than having galleries full of the crucifixion of Jesus Christ depicted in so many different ways by so many wonderful artists? \citep[original emphasis pp.~276--277]{chang2012}.
\end{quote}
Thus, the more interpretations are created the better! However, if knowledge means to learn something about the ``outer world'' and there is only one true interpretation which tells us what the world actually is like, then adding interpretations will not increase our knowledge. In fact, it will produce exactly the opposite result, the more interpretations we add the less knowledge we gain. Bueno might reply that there is no need to believe that these interpretations do in fact talk about reality (in a correspondentist sense). But then, why add interpretations in the first place? It seems we are going in circles. Differently from logical empiricists who stop at the level of empirical adequacy, constructive empiricists and their followers are willing to add not one but many ontological narratives without feeling compelled to reify the ontology, they don't have to choose nor believe in them ---which is why underdetermination is not a bug for constructive empiricists, but rather a feature. Belief is left for realist fanatics. But one might then wonder what is the meaning of such narratives if they cannot be linked to knowledge about what they refer to, namely, reality? This is keeping the narratives but detaching them from their capacity to provide knowledge. Furthermore, since the pragmatic content is already provided by the theory, it becomes unclear what such interpretations could possibly add. Interpretations become then fictional possibilities equated with not knowing and a dubious pragmatic utility. If a narrative has no underlying foundation then the fact it produces a meaningful explanation becomes just a matter of pure luck! 

There seems to exist two main possibilities. If one assumes a true link between interpretations and reality ---which neo-Pyrrhonists and constructive empiricists do not deny--- then only one theory and interpretation can be linked to reality and the addition of more and more interpretations is not a way to gain understanding but ---instead--- will lead to an extension of our ignorance (in statistical mechanics the more possibilities the more ignorant one is about the actual state of affairs). But if neo-Pyrrhonists take the other path and argue that interpretations do not provide knowledge about reality, then there seems to exist no reason, no justification in adding interpretations in the first place. Interpretations should be then considered as nothing but ``fake stories''. Now, if the claim is that made-up stories ---such as myths--- can provide an understanding equivalent to that of science we are jumping into dangerous waters. For the acceptance of made up narratives seems to blur the limit between mythical stories and scientific representations. Since neo-Pyrrhonists take that truth and understanding are essentially disconnected, a mythical story, even though false, could in principle provide understanding. According to \citet[p.~13, original emphasis]{bueno2021}: ``\textelp{} the idea is \textit{not} to provide necessary and sufficient conditions for knowledge. (It is unclear whether any such conditions can be offered anyway.)'' But then, we are left with a major difficulty for if there are no objective conditions which would allow us to make a choice between true and false interpretations why, say, the ``flat Earth interpretation'' should not be regarded as an acceptable scientific account about how things could possibly be? Why shouldn't we give credit to the anti-vaccine movement or admit that climate change deniers might be right?

\section{The withering of scientific realism}\label{sec:6}
As we have just seen the solution provided by anti-realism does not seem to work. But is there a way to get rid of underdetermination in more realist terms? In order to address this issue, let us discuss in brief \citeauthor{putnam1981}'s \citeyearpar[chap.~3]{putnam1981} distinction between ``metaphysical realism'' (MR, sometimes called ``God's Eye View'') and ``internal realism'' (IR, sometimes called ``pragmatic realism''). The former is defined as the conjunction of the three following hypotheses:
\begin{enumerate}[MR$_1$]
{\bf    \item There are only mind-independent objects in the world.
    \item There is only one true description of the world.
    \item Truth involves correspondence.}
\end{enumerate}
Of course, after Kant, supporting the claim that science can reach a description of reality-in-itself (metaphysical realism) is not just naive but simply untenable and anachronic. The idea that we can have a God's eye perspective of reality is absolutely ridiculous not only from a skeptic viewpoint but also from any meaningful realist program which attempts to relate knowledge with reality. Anti-realist are essentially correct to characterize this as ungrounded realist trend of dogmatic fanaticism. Taking this into account, Putnam proposed to restrict metaphysical realism in the following terms: 
\begin{enumerate}[{I}R$_1$]
{\bf  \item Existence questions are always relative to a theory.
    \item There is more than one true description of the world.
    \item Truth does not involve correspondence.}
\end{enumerate}
As we saw, underdetermination prevents one to justify the adoption of a scientific realist stance toward QM. The question which would naturally emerge in the theoretical level is: why (say) SQM and not Bohmian mechanics or the Dynamical Reduction Program? Then, in the ontological level, why Many Worlds and not Heisenberg's Potentiality interpretation or Popper's Propensity interpretation? If there can be only one true description (MR$_2$) of the world, which describes the world objectively as it is (MR$_1$), then we are not in a position to make such a scientific-realist \textit{pace MR} statements \textit{because of} theoretical underdetermination. One possible way out is to restrict scientific realism to internal realism \citep{ellis1988}. In this way, one person could be \textit{internally} realist about, say, Standard QM and Many Worlds interpretation and another person could be realist about Bohmian Mechanics and the Quantum Field interpretation. In this case, the theory and interpretation chosen by the internal realists will \textit{refer} only \textit{relative} to their choice. Everyone is free to choose their own reality. Now, regardless of the obvious circularity, is this realistic? Some argue that it is not.
\begin{quote}
Admittedly, Putnam's position does boast a rich ontology. Electrons exist every bit as much as chairs and tables do, and electrons can even help to \textit{explain} the superficial properties of macro-objects. Few realists, however, are willing to count this as a sufficient condition for being a ``realist.'' After all, Putnam insists that ontological commitment is always internal to a conceptual scheme; there is no scheme-independent fact of the matter about the ultimate furniture of the universe. \citep[p.~49]{anderson1992}.
\end{quote}
As \citet[p.~62]{putnam1991} recognizes, his internal realism is obviously inspired by Kant's system: ``talk of ordinary `empirical' objects is \textit{not} talk of things-in-themselves but only talk of things-for-us''. Hence, to talk about scientific theories describing the world is to be understood not as theories talking about the \textit{world-in-itself}, but as theories talking about the \textit{world-for-the-theory}. Internal realism is not more realist than Kant's empirical realism \citep{moran2000}. Internal realism is also a terminology inspired by Carnap's theory of linguistic frameworks, in which internal questions are trivial (or quickly decidable) and external questions are meaningless \citep[cf.][]{carnap1950}. For instance, consider the following statement: 
\begin{quote}
The objective chance of an outcome is the quantum weight (squared amplitude) of the set of Everett worlds in which that outcome occurs. \citep[p.~22]{wilson2020}.
\end{quote}
As Everettian worlds exist within many-worlds interpretations of Everettian quantum mechanics \citep[cf.][]{wallace2012, wilson2020}, which are defined by the axioms of the theory as ``branches'' \citep[cf.][]{everett1957, arroyoolegario2021}, the question of whether Everettian worlds exists \textit{internally} to Everettian quantum mechanics is trivially answered by \textit{yes}. But do Everettian worlds exist \textit{externally} to Everettian quantum mechanics? Put it in another way: do Everettian worlds \textit{really} exist independently of the linguistic framework in which it was defined? To Carnap, this is not even a well-formed question. Here's Putnam:
\begin{quote}
it is trivial to say that any word refers to \textit{within} the language the word belongs to, by using the word itself. What does `rabbit' refer to? Why, to rabbits, of course! What does `extraterrestrial' refer to? To extraterrestrials (if there are any). \citep[p.~52]{putnam1981}.
\end{quote}
\begin{quote}
What is wrong with the notion of objects existing ``independently'' of conceptual schemes is that there are no standards for the use of even the logical notions apart from conceptual choices. \citep[p.~114]{putnam1991}.
\end{quote}

But internal realism is not the only option through which scientific realists diminish their level of realism/commitment to what one is being a realist about. A case-in-point is what \citet[p.~5]{psillos1999} called the ``divide et impera move'' ---the basis for less-demanding realist stances, such as \citeauthor{chakravartty1998}'s \citeyearpar{chakravartty1998} ``semirealism'' \citep[cf. also][p.~142]{chakravartty2017}--- which states that scientific realists should not be committing themselves with the (approximate) truth of scientific theories \textit{as a whole}, but with \textit{parts} of itbut with \textit{parts} of it, viz. the parts worth of belief \citep[cf.][]{ruetsche2020}. In this way, as recommended by such selective realism(s), scientific realists should first distinguish between what is essential and inessential for the theory to work properly, and ---only then--- adopt a realist attitude exclusively to the former.\footnote{Such strategy was originally conceived to respond to the anti-realist historical criticism of theory change, sometimes referred to as the ``meta-pessimistic induction'' viz. how can one be a realist about scientific theories when history shows us that scientific theories change all the time? Localized dashes of realism are supposed to point out that the ``essential'' features remain unchanged through scientific changes. Even if granted that such strategy can succeed with the meta-pessimistic induction ---and even that is controversial \citep[cf.][]{tulodziecki2017}--- it is less clear, however, how such strategy responds to the argument from underdetermination which we are focusing on here.} ion between observable and unobservable comes into play, the essential parts are often the observable ones, so selective approaches to realism employ an empiricist strategy. The same goes for other so-called realist strategies, such as the ``model-dependent realism'' defended by \citet{hawking-mlodinow2010} ---which, in its turn, as pointed out by \citet[p.~360]{beebedellsen2020}, is also very similar to constructive empiricism viz. anti-realism. Take a look:
\begin{quote}
    \textit{There is no picture---or theory-independent concept of reality.} Instead we will adopt a view that we will call model-dependent realism: the idea that a physical theory or world picture is a model (generally of a mathematical nature) and a set of rules that connect the elements of the model to observations. \textelp{} According to model-dependent realism, it is pointless to ask whether a model is real, only whether it agrees with observation. \citep[pp.~42--43, p.~46, original emphasis]{hawking-mlodinow2010}.
\end{quote}

So this is where we got: scientific realism is the stance according to which scientific theories describe scientific theories. Compare that with what we wanted initially, viz. that scientific realism is the stance according to which science tells us what the world is like. If one thing, this official story on what is scientific realism teaches us is what realism is \textit{not}. From this perspective, the whole scientific realist program seems to be a red herring. Isn't there any resolution to this plot? What went wrong? In this respect \citet[sect.~1.1]{chakravartty2017sep} stresses that defining realism is easier said than done since it has been ``\textelp{} characterized differently by every author who discusses it, and this presents a challenge to anyone hoping to learn what it is.'' As we saw in the previous section, it all boiled down to a harsh tension between what scientific realism amounts to and what it actually delivers. On the one hand, realism is characterized as the claim that science describes ``the (true) furniture of the world'' or reality-in-itself (i.e., God's eye realism); on the other hand, given the difficulties for adequately justifying such a correspondentist representation of reality, scientific realism can be also relativized, either as:

\begin{itemize}
\item The claim that science describes the world ``approximately''. This can be spelled out in at least two ways.
    \begin{itemize}
    \item One can argue, as \citet{niiniluoto1987}, that the ``truthlikeness'' of successful scientific theories can be identified by a set of possible worlds in which the theory is true (in a correspondentist sense). So it may be calculated how ``close'' such possible world, in which the scientific theory in question is true, to ours ---in which its truth is approximately true \citep[see][sect.~3.4--3.5 for discussion]{niiniluoto2019}. Let us call this ``approximate realism''.
    \item Alternatively, it can be argued that there are certain ``worldly facts of the matter'' that scientific theories often stumble upon. Of course such parts of the worlds are not the whole world, nor such parts of the theories that bear such a relation of correspondence are the whole theory. Rather, they represent parts of the theory that truly corresponds to parts of the world. So one can use the partial structures formalism in order to account for that \citep[cf.][]{dacostafrench2003}. Let us call this ``partial realism''.
    \end{itemize}
\item The claim that scientific theories provide only an ``internal'' account which is dependent on the theory itself and does not \textit{refer} to the ``outer world'' (i.e., pragmatic or internal realism).
\end{itemize}
If the goal of realists is to answer Carnapian ``external questions'' par excellence, or to describe the world as-it-is or reality-in-itself, then the battle seems to be lost \citep[see][]{arroyodasilva2022}. So the situation can be depicted as follows:
\begin{quote}
For many professed `realists', realism amounts to little more than a willingness to repeat in one's philosophical moments what one says in one's scientific moments, not taking it back, explaining it away, or otherwise apologizing for it: what we say in our scientific moments is \textit{all right}, though no claim is made that it is \textit{uniquely} right, or that other intelligent beings who conceptualized the world differently from us would necessarily be getting something wrong. For many professed ``anti-realists'', realism seems rather amount to a claim that what one says to oneself in scientific moments when one tries to understand the universe corresponds to Ultimate Metaphysical Reality, that it is, so to speak, a repetition of just what God was saying to Himself when He was creating the universe. The weaker position might be called anti-anti-realism, and the stronger position capital-R Realism. \citep[p.~19, original emphasis]{burgess2004}.
\end{quote}
As Burgess stresses, the realist self-image of realism is not realism, but the negation of instrumentalism; as he calls it, it means ``anti-anti-realism''. As per the lesson learned in the last section, calling it ``realism'' doesn't make it realistic. The anti-realist image of realism is, on the other hand, what we have called MR in the previous section or ``God's-eye realism''. Is there a way to steer a middle ground between these two extremes that, while maintaining the realist's intuition of reference still copes well with the difficulties imposed by the anti-realist? It seems not. Why? Here's an idea: because the question itself seems to have already been phrased in terms that are not only disadvantageous for the realist stance but, even worse, do not even respect the main realist premises. Let us unpack this with an example in the recent literature. According to \citet[p.~5]{lamwuthrich2020}, we should withhold belief in the unobservable aspects of scientific theories which rests on ``speculative and unconfirmed status''. It is thus fairly easy to see why they don't recommend textbook scientific realism: we cannot be sure whether speculative and unconfirmed theories grasp reality.\footnote{Their example is on quantum gravity, but this can be generalized to our discussion on non-relativistic quantum mechanics.} Textbook scientific realism just wont do the job, but ---allegedly--- \textit{another} kind of realism would: \textit{presuppositional} (sic) \textit{scientific realism}.
\begin{quote}
The presuppositional version of scientific realism deems the interpretation of scientific theories a central task of philosophy (of physics). Accepting presuppositional scientific realism thus enjoins asking what the world would be like if the theory at stake \textit{were} true. \citep[p.~5, original emphasis]{lamwuthrich2020}.
\end{quote}
\citet[p.~237]{schroeren2021} also employs the above-mentioned \textit{presuppositional scientific realism}, and he defines it as ``\textelp{} the project of providing a precise account of the metaphysical ground floor in a possible world at which that theory is true''. One might call it ``realism'', yet this is a straightforwardly constructive-empiricist (i.e. anti-realist) manner of presenting the goals of science \citep[see][]{sep-constructive-empiricism}. Come on. Similar considerations can be read in what \citet{martens2022} called ``not(-yet)-realism'', which is an attitude based on epistemic humility and caution towards scientific theories which currently are facing the problem of underdetermination ---this, however (and allegedly), is not an anti-realist stance yet.

To wrap up: the present situation in scientific realism can be resumed as follows.
\begin{description}
\item[God's Eye Realism:] Scientific theories truthfully \textit{refer} to reality-in-itself.

\item[Selective Realism:] Scientific theories truthfully \textit{refer} to reality-in-itself, but only when considering parts of the theory.

\item[Approximate Realism:] Scientific theories \textit{refer} to reality-in-itself, but only approximately.

\item[Pragmatic or Internal Realism:] Scientific theories do not \textit{refer} to reality-in-itself, objects are not independent of theories, they are only to be considered either internally or pragmatically.
\end{description}
The first form of realism is almost a straw version of the realist thesis and an easy target for bullies which implies Kant's the impossibility to know anything about the \textit{noumenon}; the second divides reality into pieces of which only a few are captured by scientific theories ---and from here on the formulations represent a decrease in the realist ambitions---; the third is a relativization of \textit{truth} ---which seems an oxymoron--- that haven't yet been defined in precise terms \citep[for an attempt at this, see][]{dacostafrench2003}; and the fourth and final is a relativization ---analogous to the third--- of the \textit{referent} of theories. As it seems, \citet[p.~31]{chang2018} was indeed right when he said that the notion of ``realism'' ``\textelp{} has been hijacked by the most unrealistic of people''. However, as we argue, this is not a surprising conclusion. Fine had already hinted that the empiricist definition of scientific realism would leave realism in a dangerous maze.
\begin{quote}
In redefining realism as a doctrine about truth and belief in the truth, van Fraassen set up the debate over realism as a debate over the reach of evidence. Does the evidence support belief in the truth of our theories or does it only reach as far as belief in their empirical adequacy? Notice that this is a purely epistemological question and this is the question on which almost all the recent literature in the realism debate has centered. Still, it really is a set up. Like a skilled magician doing sleight of hand, van Fraassen's focus on the epistemological question has distracted us from what realism actually involves. Any student in a freshman philosophy course knows that realism is a metaphysical doctrine. It asserts the existence of a real, external world. In \textit{The Scientific Image} van Fraassen made that world disappear from the debate. \citep[p.~120, original emphasis]{fine2001}.
\end{quote}
Four decades later and we are still debating the empiricist account of what scientific realism is viz. leaving the world out of the scientific-realist picture. The very definition of scientific realism as a fanatic believer seems to be flawed: \textit{realistic from far but far from realism}. The present situation described so far suggests a dual diagnosis. In an optimistic analysis, scientific realism is just anti-realism, but in a critical state of denial. In a pessimistic analysis, scientific realism is a position infiltrated by anti-realism that has completely obliterated the once legitimate realist ambitions. So it seems that realism has been trapped in an anti-realist maze, but the walls are beginning to breach.

\section{Realism is dead. Why bother?}\label{sec:7}

\epigraph{``I don't want you to listen to me. I want you to listen to the scientists.''}{Greta Thunberg}

Arthur Fine described during the mid 1980s the final end of realism:
\begin{quote}
Realism is dead. Its death was announced by the neopositivists who realized that they could accept all the results of science, including all the members of the scientific zoo, and still declare that the questions raised by the existence claims of realism were mere pseudo-questions. Its death was hastened by the debates over the interpretation of quantum theory, where Bohr's nonrealist philosophy was seen to win out over Einstein's passionate realism. Its death was certified, finally, as the last two generations of physical scientists turned their backs on realism and have managed, nevertheless, to do science successfully without it. \citep[p.~112]{fine1986}.
\end{quote}
Almost four decades later, regardless of the multiplication of supposedly ``realist interpretations'', things have not changed. After all, scientific realism cannot be defended as it is. No one, not in science nor philosophy, should willingly accept being labelled as a ``fanatic believer''. For it is (objective) scientific reasons and not (subjective) personal preferences which should guide the realist quest to account for what the Ancient Greeks called \textit{physis} and was later on translated as reality. Departing from this path, while contemporary scientific realism ---as defined by anti-realists--- has been framed in truly un-scientific terms, its alternatives have ended up defending such watered-down commitments regarding their realist beliefs that it becomes hard to distance their stance from anti-realist themselves. This group of ``closet empiricists'', as Steven \citet[p.~48]{french2014} has called them, have attempted to escape \textit{truth} and \textit{reference} throwing the baby out with the bathwater. 

The situation has become the following: while the fanatic realist claims to possess ---like some sort of diviner--- a direct access to God's eye and observe reality-in-itself, the different tribes of thin realists have become solipsists who confuse reality with their ``self-made personal worlds''. This contemporary situation in science and philosophy seems to take us down a dangerous path not only for ourselves but also ---regardless of anti-realist claims--- for the world we actually inhabit. Take for instance climate change. When the young activist Greta Thunberg advised everyone to ``listen to the scientists'' at the U.S. Congress in 2021, she was referring to the \citet{ipcc2021} report on climate change which describes the irreversible human-made changes in the climate of our planet without knowing about the serious problems regarding the true \textit{reference} of such scientific report. To illustrate this point, here's Mizrahi:
\begin{quote}
\textelp{} in the case of the theory of anthropogenic climate change in climate science, the existence of processes like the transfer of infrared energy and the heat-trapping properties of greenhouse gases, such as carbon dioxide and methane, is postulated to explain global warming trends. Should we believe in the existence of these processes and entities that cannot be directly observed with the naked eye but rather are postulated in order to explain observed phenomena? \citep[p.~20]{mizrahi2020}.
\end{quote}
A similar situation was recently displayed during the COVID-19 pandemic when societies around the world begun to deny the real existence of viruses and ---consequently--- the importance of vaccines. But what other outcome could be expected if heat-trapping properties, viruses and quantum wave functions are understood as ``calculational tools'' or ``useful fictions''? If the stories scientists tell us about atoms, quantum jumps, multiverses, viruses and climate change are only narratives, ``ways of talking'' only useful for marketing theories or giving a 15min talk to the vulgus, it is not strange that our societies ---following the anti-realist trend of thought--- are becoming to observe scientific discourse from a skeptic distance. If contemporary scientists and philosophers give us the choice between accepting that reality is a meaningless notion, an impossible reference which will remain forever unknowable, or invite us to fanatically believe in their made-up narratives designed by their own personal aesthetics and preferences, why should we expect our societies to take seriously what scientists tell us about the world? After all, if realism is dead, if reality is just a fiction of the past, why should we even bother? 
\printbibliography
\end{document}